\documentstyle[12pt]{article}

\begin{document}

\thispagestyle{empty}

\begin{center}
{\Large {\bf Orbital and Spin Magnetic Dipole Strength }}
{\bf in a shell model calculation with $\Delta N$=$2$  excitations:
$^8\mbox{Be}$}

\vspace{0.3in}

M. S. Fayache and L. Zamick

\begin{small}
{\it Department of Physics and Astronomy, Rutgers University,
	Piscataway, NJ 08855}
\end{small}

\end{center}

\begin{abstract}
The magnetic dipole strength and energy-weighted strength distribution
is calculated in $^8\mbox{Be}$, as well as the separate orbit and spin
parts. All $\Delta N$=$2$ excitations over and above (and including)
the configuration $0s^4$$0p^4$ are included. The interaction
has a central, two-body spin-orbit and a tensor part. The energy-
independent and energy-weighted {\underline orbital} strength
distribution is remarkably insensitive to the presence or absence of
the spin-orbit or tensor interaction -not so the spin strength. The
energy-weighted strength distribution can be divided into a low enegy
and a high energy part. The high energy orbital part is somewhat
less but close to the low energy part, in fair agreement with a prediction
that they be equal by de Guerra and Zamick and by Nojarov. There is a
wide plateau separating the low energy part from the high energy part.
\end{abstract}

\pagebreak

\section{Introduction}
In this work we wish to study the orbit and spin magnetic dipole
strength distribution, both energy-independent and energy-weighted in
a deformed nucleus. Our interests are slanted more towards the orbital
distribution because of the intense work in recent years on scissors
mode excitations \cite{b}, \cite{ip}, \cite{ia}. The best milieux for
scissors mode excitations are strongly deformed nuclei. We pick for
our study the nucleus $^8\mbox{Be}$. The reasons for this choice are
two-fold: it is known to be strongly deformed and we can perform a
large shell model calculation which includes not only the basic
configuration $0s^4$$0p^4$ but also all $\Delta N$=$2$ excitations.
Thus we can get a low energy and a high energy strength distribution.
There are of course some atypical properties of $^8\mbox{Be}$. This
nucleus is not stable and therefore is not amenable to the most direct
way of reaching scissors modes -inelastic electron or photon scattering.
Secondly, being an $N$=$Z$ nucleus, the scissors modes will have
isospin $T$=$1$, whereas the ground state has isospin zero. The
scissors modes will be at a much higher energy than in a typical heavy
deformed nucleus where, despite the fact that the scissors modes are
isovector excitations, the $J$=$1^+$ states that have been observed
have the same isospin as the ground state.

Theoretical studies of 'scissors mode' excitations of $N$=$Z$ have
been carried out before, e.g. by Chaves and Poves \cite{cp} for
$^{20}\mbox{Ne}$.
They show that the $1^+$ $T$=$1$ states at rather high excitation
energies have all the right properties to be called scissors mode
excitations.

We emphasize again that if we were to carry out the shell model
calculation in the small model space $0s^4$$0p^4$ we would not be
adding much new to the subject. But by performing a calculation in a
larger space we are able to settle some questions about the strength
distribution. Another point we wish to pursue is how sensitive are the
energies and strength distributions to various parts of the
nucleon-nucleon interaction, in particular the spin-orbit and tensor
interactions. To this end, we use an interaction which we have used
before for other purposes -the schematic interaction of the form

\begin{equation}
V_{sche}=V_c+xV_{so}+yV_t
\end{equation}
where $c$, $s.o.$ and $t$ stand for central, spin-orbit and tensor
respectively.  The parameters of $V$, for $x$=$1$ and $y$=$1$, were
loosely fitted to the matrix elements of the Bonn A interaction. We
can vary the strength of the spin-orbit and tensor interactions by
varying $x$ and $y$.

\section{Results:}
We perform OXBASH \cite{oxbash} calculations for the $J$=$0^+$
$T$=$0$ ground state and $J$=$1^+$ $T$=$1$ states of $^8\mbox{Be}$.
We calculate the B(M1)'s from the ground state to the $1^+$ states
for three cases :

\begin{tabbing}
(1) Spin + Orbit: \hspace{.3in} \=$g_{l\pi}$=$1$ \hspace{.3in}
\=$g_{l\nu}$=$0$ \hspace{.3in} \=$g_{s\pi}$=$5.586$ \hspace{.3in}
\=$g_{s\nu}$=$-3.826$ \\
(2) Spin Only: \hspace{.3in} \>$g_{l\pi}$=$0$ \hspace{.3in}
\>$g_{l\nu}$=$0$  \hspace{.3in} \>$g_{s\pi}$=$5.586$ \hspace{.3in}
\>$g_{s\nu}$=$-3.826$ \\
(3) Orbit Only: \hspace{.3in} \>$g_{l\pi}$=$1$  \hspace{.3in}
\>$g_{l\nu}$=$0$ \hspace{.3in}  \>$g_{s\pi}$=$0$ \hspace{.3in}
\>$g_{s\nu}$=$0$ \\
\end{tabbing}

{\bf{ 2 (a) The first $7$ states}}

In Table $I$ we present results for energies and B(M1) for
the first $7$ states. We consider the three different sets of $M1$
operators as above and various choices of $x$ and $y$, the spin-orbit
and tensor strength.

What emerges from Table $I$ is that there is a well defined
scissors mode in $^8\mbox{Be}$. If we focus on the orbital column, we
see that for a pure central interaction ($x$=$0$, $y$=$0$), the
dominant state is the second one at $19.92$ $MeV$ with a strength
$B(M1)=0.61{\mu_N}^2$. It can be seen from Table $II$ that the total
orbital strength is $0.87{\mu_N}^2$ so that one gets $71\%$ of the
strength concentrated in one state when a central interaction is used.
The orbital $B(M1)$ of the $19.92$ $MeV$ state has nearly the same
value as $B(M1)$ with the full operator i.e. spin + orbit. This shows
that the mode is dominantly orbital. When the spin-orbit and tensor
interactions are turned on ($x$=$1$, $y$=$1$), there is some
fragmentation of the orbital strength. The lowest $1^+$ state, at
$17.96$ $MeV$, gets a strength of $0.398{\mu_N}^2$ and the
next one, at $20.8$ $MeV$ gets $0.198{\mu_N}^2$. The summed
strengths of these two states is about the same as in the $x$=$0$,
$y$=$0$ (central only) case. Further examination shows that this
fragmentation is mainly due to the spin-orbit interaction -the tensor
interaction is much less important. Note that the other states listed
carry negligible strength.

\vspace{.5in}

{\bf{ 2 (b) The Total Strength and Energy-Weighted Strength}}

There has been considerable interest in magnetic dipole strength $S$
and energy weighted strength $E.W.S$ distributions. First the interest
was more on the spin strength related to quenching of both isovector
magnetic dipole strength and the closely related Gamow-Teller
strength. More recently, there has been a focus on orbital strength
and energy weighted strength. This is related to the experimental
observation by W. Ziegker \cite{zr} et. al. and C. Rangacharyulu et.
al. \cite{ra} that there is a close relation between summed magnetic dipole
orbital 'scissors mode' strength and electric quadrupole strength
$B(E2)$ from the $J$=$0^+$ ground state to the first $2^+$ state in
even-even deformed nuclei.

This observation has lead to much theoretical work on energy weighted
sum rules, including works of Heyde and de Coster \cite{hd}, Zamick
and Zheng \cite{zz}, de Guerra and Zamick \cite{dz1}, Nojarov
\cite{no} and Hamamoto and Nazarewicz \cite{hz}. The above works
relate to heavy deformed nuclei where complete $\Delta N$=$2$
shell model calculations are not possible. By here considering a light
strongly deformed system $^8\mbox{Be}$, we hope to cast some light on
the theoretical works in heavier systems.

In Tables $II$, $III$ and $IV$ we present the total strength and
energy-weighted strength. We do this for the total $M1$ operator, the
spin part, and the orbital part and for various $x$ and $y$. There
are many interesting comments to be made about this table.

First of all, both the summed and energy-weighted summed {\underline
orbital} strength is remarkably insensitive to $x$ and $y$ -that is
whether the spin-orbit interaction and/or tensor interaction are
present or not present. The values of the summed strenghts for
$(x,y)$=$(0,0)$, $(0,1)$, $(1,0)$, $(1,1)$ are respectively $0.75$,
$0.74$, $0.72$ and $0.73$ ${\mu_N}^2$ while the corresponding energy
weighted numbers are $20.48$, $21.03$, $20.05$ and $20.56$
${\mu_N}^2MeV$. It is especially surprising that when a spin-orbit
splitting is introduced within the $0p$ shell, it has very little
effect. This is undoubtedly due to the fact that $^8\mbox{Be}$ is
strongly deformed, so the asymptotic wave functions are approximately
valid in all cases.

For the spin strength, there is much more sensitivity to the
interaction. It was noted by Kurath \cite{ku} that the spin-orbit
interaction is very important for magnetic dipole {\underline spin}
interactions -his energy weighted sum rule uses the spin-orbit
interaction. It was noted by Zamick, Abbas and Halemane \cite{zah} that the
tensor interaction can also have a large effect provided one allows
for ground state correlations in the nucleus.

The results in table $III$ support the claims of these authors. When
the tensor interaction is turned off $(y=0)$ then the summed spin
strength $S$ without a spin-orbit interaction is $0.12$ ${\mu_N}^2$.
When the spin-orbit interaction is turned on $S$ more than triples to
$0.38$ ${\mu_N}^2$. The corresponding energy-wighted numbers $E.W.S$
are $5.3$ ${\mu_N}^2MeV$ and $11.6$ ${\mu_N}^2MeV$.

On the other hand, with the spin-orbit interaction turned off, the
value os $S$ changes from $0.12$ ${\mu_N}^2$ to $0.40$ ${\mu_N}^2$
when the tensor interaction is turned on and $E.W.S$ increases by
about a factor of four, from $5.3$ ${\mu_N}^2MeV$ to $21.3$
${\mu_N}^2MeV$.

As compared with the isovector orbital transition, the isovector spin
transition has a factor $(9.413)^2$, which in general makes spin
transitions much larger than orbital ones. However, we see here that
the summed orbital strength is comparable -indeed somewhat larger than
the summed spin strength. This is a manifestation of the strong
deformation in $^8\mbox{Be}$. In the $SU(4)$ limit, the spin
transition rates will be zero. The asymptotic wave functions in the
$0p$ shell become $zP$, $yP$ and $xP$ where

\begin{equation}
P=N\exp(-\frac{x^2}{2{b_x}^2}-\frac{y^2}{2{b_y}^2}-\frac{z^2}{2{b_z}^2})
\end{equation}

The occupied orbits have the quanta in the $z$ direction. A transition
from '$z$' to '$x$ or '$y$' cannot be induced by the spin operator
$\sigma$.

\vspace{.5in}

{\bf{ 2 (c) The Energy-Weighted Distribution}}

We now switch from tables to figures. We show the cumulative sum of
$(E_{n}-E_{0})B(M1)$ for the total, the spin and the orbital magnetic
dipole operators in Figures 1, 2 and 3, where we consider only the
full interaction $x$=$1$, $y$=$1$. The curves for other values of
$(x,y)$ are qualitatively similar. Let us first focus on the orbital
excitations. The energy weighted sum shoots up to about $12$
${\mu_N}^2MeV$, then there is a wide plateau from about $20$ $MeV$ to
$60$ $MeV$ and then a rapid rise to $20.6$ ${\mu_N}^2MeV$ and another
plateau. Indeed there is no further change.

We can clearly identify the low-lying strength and the high-lying
strength. The low-lying energy weighted value is about $12$
${\mu_N}^2MeV$ and the total strength is $20.56$ ${\mu_N}^2MeV$. There
had been a prediction using a simple Nilsson model by de Guerra and
Zamick \cite{dz1}, \cite{dz2} that the high-lying energy weighted
strength should equal the low-lying energy weighted strength. A
similar result was obtained by Nojarov \cite{no} with a somewhat
different approach.

Our shell model calculation gives the high-lying energy weighted
strength to be $71\%$ of the low-lying strength, in fair agreement with
the previous predictions. One possible reason for a deviation is that
our single-particle splittings are not $n\hbar\omega$ but rather are
implicitly calculated in OXBASH with the schematic interaction
described in the introduction. Our single-particle splittings are
larger than $n\hbar\omega$.

Our shell model results with a $\Delta N$=$2$ truncation do not
support the claim of Hamamoto and Nazarewicz \cite{hz} that the high lying
energy weighted orbital strength should be much larger than the
corresponding low lying strength. Our results go somewaht in the other
direction.

We end by saying what we feel are the main points of interest in this
work. Firstly, from Table $IV$ we see the surprising insensitivity of
the orbital scissors mode summed strength $S$ or $E.W.S$ to the
presence or absence of the tensor and spin-orbit interactions.
Secondly, there are the shapes of the figures of cumulative $E.W.S$
strength versus excitation energy, with a wide plateau, especially in
the orbital case, separating the low lying from the high lying
strength.

\section*{Acknowledgment}
This work was supported by U.S. Department of Energy under Grant
DE-FG05-86ER-40299. We thank E. Moya de Guerra for her interest and
help. We thank M. Horoi for helping us put OXBASH on the ALPHA and for
insightful comments.

\pagebreak

\begin{small}

\end{small}

\pagebreak

\begin{small}
\noindent
\begin{center}
\samepage{
\nopagebreak{
{\bf Table I}.
The excitation energies of the seven lowest $J$=$1^+$ $T$=$1$ states
in $^8\mbox{Be}$ and the corresponding $B(M1)$ strengths for the three
cases of interest.

\begin{tabular}{|l|c|c|c|c|}\hline
Interaction & $E_{x}(1^+)$ & Total $B(M1)$ & Spin $B(M1)$ & Orbital
$B(M1)$ \\
\hline
$x=0$, $y=0$   & 18.84 & 0.6773 $10^{-4}$  & 0.7337 $10^{-4}$ & 0.1131
$10^{-6}$ \\
               & 19.92 & 0.6079            & 0.3873 $10^{-8}$ &
0.6080\\
               & 22.38 & 0.6403 $10^{-6}$  & 0.3327 $10^{-8}$ &
0.5512$10^{-6}$ \\
               & 27.06 & 0.0687            & 0.0688 & 0.2178
$10^{-7}$\\
               & 31.28 & 0.1490 $10^{-5}$ & 0.1918 $10^{-5}$ & 0.2706
$10^{-7}$\\
               & 32.47 & 0.1957 $10^{-5}$ & 0.2399 $10^{-5}$ & 0.2245
$10^{-7}$\\
               & 35.93 & 0.2732 $10^{-5}$ & 0.2751 $10^{-6}$ & 0.1273
$10^{-5}$\\
\hline\hline
$x=0$, $y=1$   & 18.72 & 0.5031 $10^{-2}$  & 0.4688 $10^{-3}$ & 0.7457
$10^{-2}$ \\
               & 19.90 & 0.5918            & 0.1015 $10^{-4}$ &
0.5873\\
               & 21.65 & 0.0201            & 0.9593 $10^{-2}$ &
0.1796 $10^{-2}$\\
               & 27.57 & 0.0677            & 0.0643 & 0.1694
$10^{-4}$\\
               & 30.97 & 0.8629 $10^{-2}$  & 0.0177 & 0.1638
$10^{-2}$\\
               & 32.42 & 0.0508            & 0.0455 & 0.1703
$10^{-3}$\\
               & 36.00 & 0.5522 $10^{-3}$  & 0.6005 $10^{-3}$ &
0.1476 $10^{-6}$\\
\hline\hline
$x=1$, $y=0$   & 18.28 & 0.6816 & 0.1103 & 0.2430 \\
               & 21.06 & 0.0577 & 0.0955 & 0.3022\\
               & 23.74 & 0.0630 & 0.2918 $10^{-2}$ & 0.0386 \\
               & 26.75 & 0.0799 & 0.0888 & 0.3485 $10^{-3}$\\
               & 33.58 & 0.4531 $10^{-3}$ & 0.9509 $10^{-4}$ & 0.1524
$10^{-3}$\\
               & 34.74 & 0.2743 $10^{-2}$ & 0.5405 $10^{-2}$ & 0.4449
$10^{-3}$\\
               & 36.75 & 0.1004 $10^{-5}$ & 0.1080 $10^{-3}$ & 0.1541
$10^{-3}$\\
\hline\hline
$x=1$, $y=1$   & 17.96 & 0.6216 & 0.1015 & 0.2208 \\
               & 20.80 & 0.1196 & 0.0397 & 0.2972\\
               & 22.89 & 0.0211 & 0.0128 & 0.0667 \\
               & 27.04 & 0.0943 & 0.0915 & 0.2214 $10^{-4}$\\
               & 33.51 & 0.8396 $10^{-2}$ & 0.0137 & 0.6453 $10^{-3}$\\
               & 34.57 & 0.0253 & 0.0249 & 0.1434 $10^{-5}$\\
               & 35.73 & 0.7579 $10^{-5}$ & 0.8102 $10^{-3}$ & 0.6613
$10^{-3}$\\
\hline\hline
\end{tabular}
}
}
\end{center}
\end{small}

\pagebreak

\begin{small}
\noindent
\begin{center}
{\bf Table II}.
The total strength $S$ of $B(M1)$ and the total energy weighted
strength $E.W.S$ in $^8\mbox{Be}$ for the four types of interaction.
\vspace{.2in}

\begin{tabular}{|l|c|c|}\hline
Interaction & $S$ & $E.W.S$ \\
\hline
$x=0$, $y=0$   & 0.86713 & 25.810 \\
\hline
$x=0$, $y=1$   & 1.1426  & 42.186 \\
\hline
$x=1$, $y=0$   & 1.0814  & 29.707 \\
\hline
$x=1$, $y=1$   & 1.2650  & 42.172 \\
\hline\hline
\end{tabular}
\end{center}
\end{small}

\begin{small}
\noindent
\begin{center}
{\bf Table III}.
The Spin strength $S$ of $B(M1)$ and the Spin energy weighted
strength $E.W.S$ in $^8\mbox{Be}$ for the four types of interaction.
\vspace{.2in}

\begin{tabular}{|l|c|c|}\hline
Interaction & $S$ & $E.W.S$ \\
\hline
$x=0$, $y=0$   & 0.12224  & 5.3329 \\
\hline
$x=0$, $y=1$   & 0.39662  & 21.267 \\
\hline
$x=1$, $y=0$   & 0.38415  & 11.604 \\
\hline
$x=1$, $y=1$   & 0.53413  & 23.126 \\
\hline\hline
\end{tabular}
\end{center}
\end{small}

\begin{small}
\noindent
\begin{center}
{\bf Table IV}.
The Orbital strength $S$ of $B(M1)$ and the Orbital energy weighted
strength $E.W.S$ in $^8\mbox{Be}$ for the four types of interaction.
\vspace{.2in}

\begin{tabular}{|l|c|c|}\hline
Interaction & $S$ & $E.W.S$ \\
\hline
$x=0$, $y=0$   & 0.7451  & 20.479 \\
\hline
$x=0$, $y=1$   & 0.7426  & 21.032 \\
\hline
$x=1$, $y=0$   & 0.7196  & 20.005 \\
\hline
$x=1$, $y=1$   & 0.7261  & 20.565 \\
\hline\hline
\end{tabular}
\end{center}
\end{small}

\end{document}